\documentstyle[12pt]{article}

\def\tr{\mathop{\rm tr}}
\def\spec{\mathop{\rm spec}}

\def\P{U_\Pi}
\def\O{U_\Omega}
\def\U{{\cal U}}
\begin{document}

\begin{titlepage}
\begin{flushright}
SPbU-IP-96-10\\
hep-ph/9605402
\end{flushright}
\vskip 1.5in
\begin{center}
{\LARGE The Extended Chiral Bosonisation And\\[12pt]
Pion-Diquark Effective Action}
\vskip 1in
{\large  A. G. Pronko }
\vskip 12pt
{Theoretical Physics Department, St Petersburg State University,\\
Ulyanovaskaya st 1, 198904 St Petersburg, Russia\\[4pt]
e-mail:~{\tt agp@kom.usr.pu.ru} }
\end{center}
\vskip .5in

\begin{abstract}
\normalsize

We consider bosonisation of the low-energy QCD
based on integrating the anomaly of the
extended chiral (E$\chi$) transformation which depends both on
pseudoscalar meson and scalar diquark fields as parameters.  The
relationship between extended chiral and usual chiral anomalies and related
anomalous actions is studied.  The effective action for the extended chiral
field $\U$ depending on complete set of anomalous generators of the
E$\chi$-transformation is given. The terms of this effective action
relevant to interaction of pions and $\bar 3_c$ scalar diquarks are written
down explicitly.

\end{abstract}
\end{titlepage}

\section{Introduction}

In a conventional understanding of the QCD physics diquarks [1,2]
and the quantum anomalies [3] usually are considered as unrelated topics.
However in the recent paper [4]  a possible
connection between these two areas has been proposed.
In [4] it has been shown that $J^P=0^+$ diquarks can be introduced
together with pseudoscalar mesons
as parameters of certain transformations --- extended chiral (E$\chi$)
transformations, including usual chiral as a subgroup.
The chiral bosonisation method based on integrating
chiral anomaly has been applied in
order to obtain the effective action for diquarks and to study their
properties.  The values for the mass of the $ud$-diquark $m_{ud}\approx 300
MeV$ and mean square charge radius $\langle r^2_{ud}\rangle^{1/2}\approx
0.5 fm$ obtained in ref.[4] stimulate an interest for further
investigations in this direction.

The present paper is devoted to calculation of the effective interaction of
pions and scalar diquarks within the approach based on
E$\chi$-transformations. This is a natural extension of the results of the
ref.[4] on the case when the extended chiral field $\U$ depends on
pions, scalar diquarks and other fields needed for closure of the
E$\chi$-group [4]. All these states are introduced on a quasi-goldstone
manner as angle-like variables corresponding to anomalous (non-preserving
measure of quark path integral) generators of E$\chi$-transformations.
An especial status of pions which are as usually supposed to be goldstone
particles in the chiral limit is regarded by means of
choosing  the parameterization of the field $\U$ in a certain form.
All results of the pion physics known from the chiral bosonisation approach
[5] are reproduced. We do not introduce any new parameters.

The Sect.2 contains main elements of the chiral bosonisation approach
adopted to our purposes. At first, we remind the structure of the
E$\chi$-transformations and consider the symmetry properties of the quark
path integral. After that, the E$\chi$-bosonisation is formulated.
At last, herein we study the correspondence between the extended
chiral and usual chiral anomalies and related anomalous actions.

The Sect.3 is devoted to studying the effective action $W_{eff}(\U)$
which contains information about low-energy pion-diquark interaction.
We start from the explicit expression for $W_{eff}$ in terms of the field
$\U$. The $\bar 3_c$ scalar diquark sector of $W_{eff}$ is studied and
main properties of the diquark are exhibited. Next, we discuss the proper
parameterization of the field $\U$ which regards the goldstone nature of
pions. We finish the Sect.3 by resulting expression for the effective
interaction lagrangian of pions and $\bar 3_c$ scalar diquarks.

\section{The E$\chi$-bosonisation of the quark path integral}

\subsection{The E$\chi$-transformations of the quark fields}

Let us remind the
formulation of the E$\chi$-transformations following ref.[4].
Introduce the Majorana-like eight-component spinor $\Psi$ constructed
from the ordinary Dirac spinor $\psi$
\begin{equation} \label{Psi}
\Psi=
\left( \begin{array}{c}
\psi \\
C\bar\psi^T
\end{array} \right),\qquad
\overline{\Psi}^T=
\left(\begin{array}{cc}
0  & C^{-1}\\
C  & 0
\end{array}\right) \Psi
\end{equation}
where $C$ is the charge conjugation matrix. The second relation of
(\ref{Psi}) is the Majorana condition [7].
The E$\chi$-transformations of the fermion field
with $N =N_c\times N_f$ internal degrees of freedom, are [4]
\begin{equation} \label{Phi-Theta}
\delta \Psi =\left[ \Phi+\Theta\gamma_5\right] \Psi,\qquad
\Phi  =  i\left(\begin{array}{cc}
\alpha   &  \beta  \\
-\beta^*   &  -\alpha^T
\end{array}\right),
\quad
\Theta=i\left(\begin{array}{cc}
\chi      & \omega \\
\omega^* & \chi^T
\end{array}\right).
\end{equation}
The matrices $\alpha,\chi,\beta,\omega$ acting in the
direct product of colour and flavour spaces
are supposed to satisfy the following hermiticity and
symmetry properties
\begin{equation} \label{acxo}
\alpha=\alpha^+ ,\quad
\chi=\chi^+,\quad
\beta=-\beta^T,\quad
\omega=\omega^T.
\end{equation}
The generators $\omega$ are of quantum numbers of scalar diquarks.
The colourless part of $\chi$ corresponds to pions.
In ref.[4] it has been shown that E$\chi$-group is $G=U(2N)$, the $\Phi$
generators span the Lie algebra of $H=O(2N)$ and the $\gamma_5$ generators
$\Theta$ belong to the coset $G/H=U(2N)/O(2N)$.
The transformations (\ref{Phi-Theta}) are most general global
transformations under whose action the kinetic term of the fermion field is
invariant.

To study the quantum theory let us introduce external vector
fields $v_\mu$, $a_\mu$, $\phi_{5\mu}$, $\phi_\mu$ generating
$\bar\psi\psi$ and $\psi\psi$ vector currents and
which have the same matrix structure as the
generators (\ref{acxo}) respectively.
In terms of spinor $\Psi$ the quark lagrangian can be written as
\begin{equation} \label{lag}
{\cal L}_\psi = \frac{1}{2}\ \overline{\Psi}\ \hat G\ \Psi,\qquad
\hat G = i \gamma^\mu (\partial_\mu+V_\mu+\gamma_5 A_\mu)
\end{equation}
where antihermitian fields
$V_\mu$ and $A_\mu$ have the block structure
\begin{equation} \label{VA}
V_\mu = \left(
\begin{array}{cc}
v_\mu &  \phi_{5\mu} \\
\phi^*_{5\mu}  & -v^T_\mu
\end{array}
\right),
\qquad
A_\mu=\left(
\begin{array}{cc}
a_\mu &  \phi_{\mu} \\
-\phi^*_{\mu}  & a^T_\mu
\end{array}
\right).
\end{equation}
Any local transformation of the fermion field $\Psi$ can be
represented equivalently as a transformation of the fields $V_\mu,A_\mu$.
The finite $\gamma_5$-transformation of the fields $V_\mu,A_\mu$
is ($U_\Theta\equiv\exp\Theta$)
\begin{eqnarray} \label{V'A'}
V_\mu &\to& V_\mu^{\Theta}=\frac{1}{2}\left\{
U_\Theta^{-1}(\partial_\mu+V_\mu+A_\mu)U_\Theta
+U_\Theta(\partial_\mu+V_\mu-A_\mu)U_\Theta^{-1}\right\}
\nonumber\\
A_\mu &\to& A_\mu^{\Theta}=\frac{1}{2}\left\{
U_\Theta^{-1}(\partial_\mu+V_\mu+A_\mu)U_\Theta
-U_\Theta(\partial_\mu+V_\mu-A_\mu)U_\Theta^{-1}\right\}.
\end{eqnarray}

The quark path integral --- generating functional for the $\bar\psi\psi$ and
$\psi\psi$ ($\bar\psi\bar\psi$) vector currents reads [9]
\begin{equation} \label{Z-psi}
Z_\psi = \int {\cal D}\Psi\
\exp \left\{ i\int d^4x\  {\cal L}_\psi \right\}
= \left[\det\hat G \right]^{1/2}.
\end{equation}
Because of the noninvariance of the measure
with respect to $\gamma_5$-transformations [10]
the quark path integral (\ref{Z-psi}) does not invariant under action of
transformations (\ref{Phi-Theta}) or (\ref{V'A'}).
Thus $Z_\psi$ has the structure $Z_\psi=\exp i(W_{an}+W_{inv})$
where $W_{an}$ and $W_{inv}$ are noninvariant and invariant functionals
of the external fields correspondingly.
This means that there exists an E$\chi$-anomaly, which is defined as follows
\begin{equation} \label{anomaly}
{\cal A}^a = \left.
\frac{1}{i}\frac{\delta\ln Z_\psi}{\delta\Theta_a} \right|_{\Theta_a=0}
=\left.\frac{\delta W_{an}}{\delta\Theta_a} \right|_{\Theta_a=0}
\end{equation}
where $\{\Theta\}=\{\chi,\omega,\omega^*\}$ is the set of
parameters of anomalous transformations; $\Theta=\Theta^a T^a$ and
$T^a$ are (block) antihermitian generators.

\subsection{The E$\chi$-bosonisation}

In refs.[6] it has been shown that if the quark path integral has an
anomaly with respect to some transformations then we can perform a
bosonisation of the variables generating these anomalous transformations.
This general result plays an important role in our suggestion to
consider scalar diquarks as associated with $\omega$ while pseudoscalar
mesons as usually will be associated with $\chi$.

Let us now formulate the E$\chi$-bosonisation.
This can be done by generalization of the chiral bosonisation approach
developed in refs.[5,6].

An important ingredient of the
chiral bosonisation of the low-energy QCD is the hypothesis that there
exists some low-energy region $L$ where the noninvariant or anomalous
fluctuations of quark variables are essential while out of this region
they are negligible. The region $L$ can be formulated as a some region of
the spectra the Dirac operator $\hat G$, as it done below. Because $Z_\psi$
is the determinant, $Z_\psi={\det}^{1/2}\hat G\equiv\prod\lambda^{1/2}$,
$\lambda\in\spec\hat G$, the hypothesis of existing the low-energy region $L$
means that $Z_\psi=Z_\psi^L\ {\widetilde Z}_{inv}$. The functional
$Z_\psi^L=\prod_{\lambda\in L}\lambda^{1/2}$ contains all information on
$\Theta$-noninvariant processes while ${\widetilde Z}_{inv}$ is invariant
functional. Thus all contributions in E$\chi$-anomaly comes from $Z_\psi^L$.

The E$\chi$-bosonisation can be formulated as follows
\begin{eqnarray} %\label{}
Z_\psi^L(V,A)&=&Z_{inv}\int {\cal D}\mu(\Theta)
\left[\frac{Z_\psi^L(V,A)}{Z_\psi^L(V^{\Theta},A^{\Theta})}\right]
\nonumber\\
&=& Z_{inv}\int {\cal D}\mu(\Theta)
\exp\left\{i W_{eff}(\Theta)\right\}
\end{eqnarray}
where ${\cal D}\mu(\Theta)$ is an invariant measure on the
corresponding coset space $G/H$ and $Z_{inv}$ is some invariant
functional. The effective action $W_{eff}(\Theta)$ is defined as
\begin{equation} \label{W-W}
W_{eff}(\Theta)= W_{an}(V,A)-W_{an}(V^{\Theta},A^{\Theta})
= - \int_{}^{} d^4 x \,\int_{0}^{1} ds \,
{\cal A}^a (V^{s\Theta},A^{s\Theta}) \Theta^a.
\end{equation}
where argument $s\Theta$ means corresponding transformation of background
fields. The effective action $W_{eff}$ is the main object which we
are looking for. It describes the effective low-energy interaction
of the $\Theta$-variables.
The explicit expression for $W_{eff}$
is written in the next Section, where it is studied.

\subsection{The E$\chi$-anomaly and $W_{an}$}

In (\ref{W-W}) the action $W_{eff}$ is defined through E$\chi$-anomaly or,
equivalently, through $W_{an}$. Thus we have to find the explicit
expressions for ${\cal A}^a$ or $W_{an}$ in terms of the external fields.

To find the E$\chi$-anomaly and the related
anomaly action one should regularize the functional $Z_{\psi}$.
Usually the chiral bosonisation method is armed by
finite-mode regularization scheme [11, 5] which consists in continuation to
Euclidian domain and specifying the low-energy region $L$ as some region of
spectra of the hermitian operator $\hat G_E$. The functional $Z_\psi^L$ in
Euclidian space is
\begin{eqnarray} \label{Z^L}
Z_{\psi,E}^L = \det{}^{1/2}
\left\{
\hat G_E \,\theta\biggl(1-\frac{(\hat G_E-M)^2}{\Lambda^2}\biggr)
\right\},\quad
\theta(A)=\int\limits_{-\infty}^{\infty}\frac{ds\,\exp(isA)}{2\pi i(s-i0)}.
\end{eqnarray}
where $\Lambda,M$ are parameters defining the region $L$: $\lambda\in L$ if
$-\Lambda+M\leq\lambda\leq\Lambda+M$;
$\lambda$ is an eigenvalue of $\hat G_E$.

The straightforward calculations based on the
regularized functional (\ref{Z^L}) lead us to the following simple but
useful result.

In fact, the expression for the E$\chi$-anomaly ${\cal A}^a$ can
be obtained from the usual chiral's one, expressed in terms of
$v_\mu,a_\mu$ only.  Indeed, in our case the operator $\hat G$ is presented
in the form of the usual Dirac operator (see (\ref{lag})) where instead of
$v_\mu,a_\mu$ we have block antihermitian external fields $V_\mu$ and
$A_\mu$ given by (\ref{VA}).  Because $V_\mu, A_\mu $ are
transformed infinitesimally by $\Theta$ like $v_\mu, a_\mu$
are transformed by usual chiral
$\chi$, the $E\chi$-anomaly  can be obtained from the known chiral one
[3,5,11] by means of formal substitution
\begin{equation} %\label{}
v_\mu \to V_\mu, \quad
a_\mu \to A_\mu,  \quad
t^a \to T^a,
\end{equation}
accomplished by taking additionally trace
in two-dimensional ``block'' space
and by taking into account the overall factor 1/2 coming from the power of
determinant in (\ref{Z^L}).  The same arguments are valid when $\hat G$
depends on external scalar fields as well.

As it is well known [3,5], the anomaly consists out of topologically trivial
${\cal A}^+$ and non-trivial ${\cal A}^-$ parts:
${\cal A}={\cal A}^+ +{\cal A}^-$.
The part ${\cal A}^-$ after integration in (\ref{W-W}) form the
Wess-Zumino-Witten action responsible for ${\cal P}$-odd processes.
The dynamics (decay constants,
masses ets.) is contained in that part of $W_{eff}$ which is
related to ${\cal A}^+$
and this part is relevant for our investigation. Instead of
${\cal A}^+$ we present here the corresponding part of $W_{an}$
\begin{eqnarray} \label{Wan}
W_{an}^+&=&\int d^4x \biggl\{
\frac{\Lambda^2-M^2}{8\pi^2} \tr_{(b,c,f)} A_\mu^2
\nonumber\\ &&
-\frac{1}{96\pi^2} \tr_{(b,c,f)}
\left(
F_{\mu\nu}^2  - [A_\mu,A_\nu]^2 - 2F_{\mu\nu}[A_\mu,A_\nu]
-2 [D_\mu,A_\mu]^2 +4 (A_\mu^2)^2
\right)
\biggr\}.
\end{eqnarray}
Herein $F_{\mu\nu}=\partial_\mu V_\nu-\partial_\nu V_\mu+[V_\mu,V_\nu]$ and
$D_\mu=\partial_\mu+V_\mu$.

Note that the arguments presented above about substitution of the fields
are valid for the expression of $W_{an}$ as well.
Therefore the results of ref.[5] can be used straightforwardly
in order to obtain the
expressions both for $W_{an}$ and ${\cal A}^a$ if it is needed to take into
account the scalar fields together with the vector ones.

\section{Pion-diquark effective action}

\subsection{An expression for the $W_{eff}$}

Due to relationship between E$\chi$-anomaly and chiral anomaly
established in previous Section, the pion-diquark effective action
in terms of E$\chi$-field $\U$ looks like as pure pion one [5] and reads
\begin{eqnarray} \label{Weff}
W_{eff}&=&\int d^4 x \biggl\{
\frac{f_\pi^2}{48}\tr_{(b,c,f)} (D_\mu \U) (D^\mu \U^+)+
\frac{1}{192\pi^2}\tr_{(b,c,f)}\biggl((D_\mu^2 \U)(D_\nu^2 \U^+)+
\nonumber\\ &&
+\frac{1}{2}(D_\mu \U)(D_\nu \U^+)(D^\mu \U)(D^\nu \U^+)
-\left[(D_\mu \U)(D^\mu \U^+)\right]^2 +
\nonumber\\ &&
+2(D_\mu F^{\mu\nu})\left[(D_\nu \U)\U^+ +(D_\nu \U^+)\U\right]
-\frac{1}{2}[F_{\mu\nu}, \U] [F^{\mu\nu}, \U^+]\biggr).
\biggr\}
\end{eqnarray}
Herein we drop the Wess-Zumino-Witten action.
For sake of simplicity
we also drop all external fields except the vector field $v_\mu$
containing relevant dynamical fields of gluons
$v_\mu=G_\mu=gG_\mu^a({\lambda^a\over 2i})$.
{}From now on
$V_\mu$ and its strength tensor $F_{\mu\nu}$ are block-diagonal matrices
\begin{equation} \label{VF}
V_\mu=
\left(
\begin{array}{cc}
G_\mu & 0 \\
0  & -G_\mu^T
\end{array}
\right),\quad
F_{\mu\nu}=\partial_\mu V_\nu - \partial_\nu V_\mu +[V_\mu,V_\nu]
=\left(
\begin{array}{cc}
G_{\mu\nu}& 0 \\
0  & -G_{\mu\nu}^T
\end{array}
\right).
\end{equation}
The covariant derivative acts as $(D_\mu *)=(\partial_\mu
*)+[V_\mu,*]$ with $V_\mu$ given by (\ref{VF}).

The parameters $\Lambda$ and $M$
are related to pion decay constant $f_\pi=132 MeV$
\begin{equation} %\label{}
f_\pi^2=\frac{3}{2\pi^2}(\Lambda^2-M^2).
\end{equation}
This relation  has been used in (\ref{Weff}) to express the only parameter
of $W_{eff}$ in a model independent way.

The E$\chi$-field $\U$ in (\ref{Weff}) is defined as
$\U=U_\Theta^2$ where $U_\Theta=\exp\Theta$ as above. But below we
{\it redefine}  $\U$. Before explanation of the reasons of
redefinition $\U$ we would like to consider the diquark sector of the
effective action (\ref{Weff}).

\subsection{The $\bar 3_c$ scalar diquark sector of $W_{eff}$}

The $\bar 3_c$ scalar diquark sector of $W_{eff}$
corresponds to the case when
$\Theta$  contains $\bar 3_c$ scalar diquark fields {\it only}, i.e.
\begin{equation} \label{ww}
\Theta=
i\left( \begin{array}{cc}
0 & \omega \\
\omega^* & 0
\end{array} \right),\quad
(\omega)^{ab}_{jk}
=\frac{1}{f_\omega}\,\omega_c\, (i\sigma_2)_{jk}\,\epsilon^{cab},\quad
(\omega^*)^{ab}_{jk}
=\frac{1}{f_\omega}\,\omega^*_c\, (i\sigma_2)_{jk}\,\epsilon^{cab}
\end{equation}
where $j,k$ are flavour and $a,b,c$ are colour indices.
$f_\omega$ is the diquark decay constant which will be
determined below.

The reduction (\ref{ww}) has been considered in [4]. It has been shown
that in this case $G=SU(4)\sim O(6)$, $H=SU(3)\times U(1)$ and
scalar $\bar 3_c$ $ud$-diquarks $\omega$ belong to
$G/H=CP^3=SU(4)/SU(3)\times U(1)$.

The diquark mass term
comes from the last term of (\ref{Weff}), because the later
contains the term $G_{\mu\nu}^a G^{\mu\nu,a}\omega^{}_c\omega^*_c$.
Substituting quasi-classically instead of $G_{\mu\nu}^a G^{\mu\nu,a}$
its vacuum expectation value we obtain the following expression for
the inverse diquark propagator
\begin{equation} \label{prop}
[{\cal D}_\omega(p^2)]^{-1}
=\frac{2f_\pi^2}{3 f_\omega^2} p^2
+ \frac{1}{12 \pi^2 f_\omega^2} p^4 -\frac{C_g}{36 f_\omega^2}
\end{equation}
where $C_g$ is the gluon condensate,
$C_g=<\frac{g^2}{4\pi^2}(G_{\mu\nu}^c)^2>$. The second term in
(\ref{prop}) comes from the ``tachionic'' term in (\ref{Weff}).
{}From (\ref{prop}) we see that the mass of $ud$-diquark
$\omega$ is defined by the gluon condensate
\begin{equation} \label{M-ud}
M_\omega^2=
2\pi^2 f_\pi^2\left(\sqrt{1+\frac{C_g}{12\pi^2 f_\pi^4}}-1\right).
\end{equation}
For actual value of gluon condensate $C_g=(365 MeV)^4$ we get
$M_\omega\approx 300 MeV$. As it was mentioned in ref.[4]
the correction of this evaluation due to quark
masses is provided by $M_\omega^2(m_q\ne 0)=M_\omega^2(m_q=0)+m_\pi^2$.
This gives $M_\omega^2(m_q\ne 0)\approx 340 MeV$, which falls into the
region allowed in the other models [2]. The low mass of diquark
can be explained by a reason that this is in fact ``current'' mass
because we do not take into account the interaction with gluons
which leads to ``dressing'' the diquark.

The diquark decay constant $f_\omega$ is defined by requirement that the
residue of the diquark propagator at $p^2=M_\omega^2$ is unity,
\begin{equation} \label{Fdi}
f_\omega^2=\frac{2}{3}f_\pi^2+
\frac{1}{6\pi^2}M_\omega^2= \frac{2}{3}\sqrt{f_\pi^4+\frac{C_g}{12\pi^4}}.
\end{equation}
For the values $C_g=(365 MeV)^4$ and $f_\pi=132MeV$ this gives
$f_\omega\approx 120MeV$. The additional term depended upon the diquark
mass in (\ref{Fdi}) comes due to tachion pole of the propagator
(\ref{prop}) and is about 5\%
respect to the first term therein. In rough
estimations it can be neglected.

In fact, the relation $f_\omega=\sqrt{\frac{2}{3}}f_\pi$ reproduce the
results of the ref.[12] where instead of $f_{\pi,\omega}$ therein
$g_{\pi,\omega}$
has been used\footnote{In [12] $g_{\omega}$ is denoted by $g^D_3$}.
In the chiral limit which we consider
here, $f_{\pi,\omega}$ and $g_{\pi,\omega}$ are related through quark
condensate $C_q(<0)$ as
\begin{equation} \label{gg}
g_\pi=-\frac{C_q}{2f_\pi},\quad
g_\omega=-\sqrt{\frac{2}{3}}\frac{C_q}{2f_\omega}.
\end{equation}
{}From $f_\omega=\sqrt{\frac{2}{3}}f_\pi$ follows that $g_\pi=g_\omega$
what is the result of ref.[12] obtained by QCD sum rules technique.

\subsection{The E$\chi$-field $\U$ with all possible $\Theta$-variables}

Consider now the most general case when E$\chi$-field $\U$ depends on
all possible $\Theta$-variables.
We take $\Theta$ in the general form (\ref{Phi-Theta})
\begin{equation} %\label{}
\Theta=i\left(\begin{array}{cc}
\chi      & \omega \\
\omega^* & \chi^T
\end{array}\right).
\end{equation}
The
$J^P=0^-$ meson states are contained in $\chi$ and $J^P=0^+$ diquark states
are contained in $\omega$
\begin{equation} \label{1836}
\chi=\frac{\pi}{f_\pi} + \frac{\chi_8}{f_8},\quad
\omega=\frac{\omega_{\bar 3}}{f_{\bar 3}} +\frac{\omega_6}{f_6}.
\end{equation}
The $SU(3)$ colour group representations of the fields are shown in
notations explicitly, except for the colour singlet fields describing
pions
\begin{equation} \label{pi}
\pi = \pi^i (\frac{\sigma^i_f}{\sqrt{2}}\otimes 1_c)
=\left(
\begin{array}{cc}
\frac{\pi^0}{\sqrt{2}}  & \pi^+\\
\pi^- & -\frac{\pi^0}{\sqrt{2}}
\end{array}
\right)_f\otimes 1_c,
\end{equation}
where $\sigma^i_f$ are isospin Pauli matrices and $\pi^i$, $i=1,2,3$ are
elements of the isotriplet, related to $\pi^{\pm,0}$ via (\ref{pi}).
Due to exceptional role of pions it is natural to extract them from the
other states in $\Theta$. Denote
\begin{equation} \label{Pi-Omega}
\Pi = i
\left(\begin{array}{cc}
\frac{\pi}{f_\pi} & 0\\[6pt]
0 & \frac{\pi^T}{f_\pi}
\end{array}\right),\quad
\Omega = i
\left(\begin{array}{cc}
\frac{\chi_8}{f_8} &
\frac{\omega_{\bar 3}}{f_{\bar 3}} +\frac{\omega_6}{f_6}\\[6pt]
\frac{\omega^*_{\bar 3}}{f_{\bar 3}} +\frac{\omega^*_6}{f_6} &
\frac{\chi_8^T}{f_8}.
\end{array}\right)
\end{equation}
Then the sum $\Theta =\Pi + \Omega$ represents decomposition on the colour
singlet ($\Pi$) and non-singlet ($\Omega$) parts.

Now we are going to discuss the reasons for redefinition
of the $\U$ if it is taken in the usual form $\U=\exp 2\Theta$. Also we
describe the most obvious method of its redefinition which means
changing variables from unphysical to physical ones.

On this way we take in advance the {\it goldstone} nature of pions
which are related to exact $SU(2)_L\times SU(2)_R$ symmetry of the
massless QCD lagrangian. The goldstone particles have an important and
general property: they can interact between themselves and
with any other particle only through vertices with derivatives.
As far as pions are considered as goldstone particles
this property must be preserved in description of those physical processes
in whose pions take part [8].

However this property will be lost if one identifies $\U$ in a usual way
as $\U=\exp 2\Theta =\exp 2(\Pi+\Omega)$ where both $\Pi$ and $\Omega$
are put to be nonzero. The puzzle arises due to the term
$\tr[F_{\mu\nu}, \U] [F^{\mu\nu}, \U^+]$ in $W_{eff}$ which contains
pion interaction terms starting from the $\Theta^4$. This happens only
because of specific dependence on pion fields in $\U=\exp 2(\Pi+\Omega)$
with $\Pi$ and $\Omega$ are given by (\ref{Pi-Omega}).
Because $[\Pi,F_{\mu\nu}]=0$, there are no pure pion vertices (together with
$G_{\mu\nu}G_{\mu\nu}$) in $\tr[F_{\mu\nu}, \U] [F^{\mu\nu}, \U^+]$.
But because $[\Pi,\Omega]\ne 0$, we find that the term
$\tr[F_{\mu\nu}, \U] [F^{\mu\nu}, \U^+]$ contains
$\pi$--$\omega_{\bar 3},\omega_6,\chi_8$ interaction terms without
derivatives. All this indicates that $\U$ taken in the form
does not preserve the goldstone nature of pions in the
$\pi$--$\Omega$ interaction sector of $W_{eff}$.

Therefore when the term  $\tr[F_{\mu\nu}, \U] [F^{\mu\nu}, \U^+]$ (and only
this term in $W_{eff}$) contains non-goldstone type of pion vertices,
we have to redefine the field $\U$ in order to eliminate the dependence of
$\tr[F_{\mu\nu}, \U] [F^{\mu\nu}, \U^+]$ on pion fields. This means that
instead of $\U=\exp 2(\Pi+\Omega)$ we have to take some
$\U_{phys}=\U_{phys}(\Pi,\Omega)$ which must satisfy the following
condition
\begin{equation} \label{mycond}
\frac{\delta}{\delta \pi^i}
\tr [F_{\mu\nu}, \U_{phys}] [F^{\mu\nu}, \U_{phys}^+] =0
\end{equation}
provided $\U|_{\Omega=0}=\U_{phys}|_{\Omega=0}$ keeps the pion sector
of $W_{eff}$ unchanged.

To determine $\U_{phys}$ we can use the relationship between the finite
$\gamma_5$-transformation and the corresponding E$\chi$-field. Choosing
any specific form of parameterization of the finite
$\gamma_5$-transformation we get some specific form of the E$\chi$-field
trying to satisfy the condition (\ref{mycond}). The following
parameterization of the element $g\in G/H$ is suitable
for our purposes
\begin{equation} \label{g}
\quad g=\exp\Pi\gamma_5\,
\exp\Omega\gamma_5
=\P \O \left(\frac{1+\gamma_5}{2}\right)
+\P^{-1} \O^{-1} \left(\frac{1-\gamma_5}{2}\right)
\end{equation}
where $\P=\exp\Pi,\, \O=\exp\Omega$.
The corresponding E$\chi$-field is $\U_{phys}$
\begin{equation} \label{Uphys}
\U_{phys}=\P \,\O \,(\P^{-1}\,\O^{-1})^{-1}=\P\,\O^2\,\P
\end{equation}
because due to $[\Pi,F_{\mu\nu}]=0$ from (\ref{Uphys})
immediately follows
\begin{equation} \label{t=t}
\tr_{(b,c,f)}[F_{\mu\nu}, \U_{phys}] [F^{\mu\nu}, \U_{phys}^+] =
\tr_{(b,c,f)}[F_{\mu\nu}, \O^2] [F^{\mu\nu}, \O^{-2}]
\end{equation}
and therefore the condition (\ref{mycond}) is satisfied.

Note that reparameterization of the E$\chi$-field does not affect the
properties of the states (\ref{1836}) entering in the $\Theta$.
It only corrects the structure of the $\pi$--$\Omega$ interaction and
does not touch  any term from the $\Omega$-sector of the $W_{eff}$.
This occurs because the reparameterization of the element $g\in G/H$ does not
change the $W_{eff}$ considered as a function of the field $\U$. The
functional dependence of $W_{eff}$ on $\U$ is closely determined by
the anomaly action $W_{an}$ whose expression in terms of external fields
 (\ref{Wan}) obviously is
independent on parameterization of the finite $\gamma_5$-transformation.

The described scheme can be easily extended on the case of three flavours.
The presence of the additional terms due to nonzero quark masses does not
influence the arguments just explained. The condition
(\ref{mycond}) is the only condition which has to be satisfied in order to
guarantee that in the chiral limit the pseudoscalar colourless mesons
appear to be exact goldstones.

\subsection{Pion-diquark effective interaction}

In what follows we will assume that $\U$ in $W_{eff}$ always is
$\U_{phys}$ given by (\ref{Uphys}). Introduce the following
notations
\begin{equation} %\label{}
\U_\Omega = \O^2,\quad
\Xi=\P=
\left(\begin{array}{cc}
\xi & 0\\
0 & \xi^T
\end{array}\right)
\end{equation}
where as conventionally $\xi=\exp(i\pi/f_\pi)$ and the pion matrix $\pi$ is
given by (\ref{pi}). In these notations E$\chi$-field
$\U$ acquires the form
\begin{equation} \label{XUX}
\U= \Xi\, \U_\Omega\, \Xi.
\end{equation}

Substituting $\U$ given by (\ref{XUX}) into (\ref{Weff}) we obtain the
following expression for $\pi$--$\Omega$ interaction lagrangian
\begin{eqnarray} \label{LpO}
L_{\pi,\Omega} &=& \frac{f_\pi^2}{48}
\tr_{(b,c,f)}\Bigl\{
[\Xi^+(\partial_\mu \Xi),\U_\Omega^{}][\Xi(\partial_\mu \Xi^+),\U_\Omega^+]
\nonumber\\  &&
+2 \U_\Omega^+ (D_\mu \U_\Omega^{}) \,\Xi(\partial_\mu \Xi^+)
+2 \U_\Omega^{} (D_\mu \U_\Omega^+) \,\Xi^+(\partial_\mu \Xi)
\Bigr\}
+ L_{\pi,\Omega}^{(4)}
\end{eqnarray}
where $L_{\pi,\Omega}^{(4)}$ contains vertices of order $p^4$.
The expression for $L_{\pi,\Omega}^{(4)}$ is rather long and
it will be presented elsewhere.
The terms of (\ref{LpO}) shown explicitly determine the main
contributions in the vertex Green functions at small external momenta.
Here we would like to stress again that the term
$\tr[F_{\mu\nu}, \U] [F^{\mu\nu}, \U^+]$ of $W_{eff}$  does not
contribute in $L_{\pi,\Omega}$ (see Sect.3.3). This term
is closely contained in $\Omega$-sector of $W_{eff}$ which can be found
if we put $\U=\U_\Omega$ in (\ref{Weff}).

The interaction of pions and $\bar 3$ scalar diquarks can be obtained
from (\ref{LpO}). Using (\ref{ww}) we find
\begin{equation} \label{pi-di}
L_{\pi,\bar 3} =
-\frac{f_\pi^2}{12}\, \sin^2|\omega_{\bar 3}|\,
\tr_{(f)}\left\{\partial_\mu u^+ \partial_\mu u \right\}
+ L_{\pi,\bar 3}^{(4)}
\end{equation}
where
\begin{equation} %\label{}
u=\xi^2=\exp(2i\pi/f_\pi),\quad
|\omega_{\bar 3}|\equiv
2\sqrt{\omega^{a}_{\bar 3}\omega^{*a}_{\bar 3}}/f_\omega.
\end{equation}

The expression (\ref{pi-di}) provides us an information about
pion-diquark effective interaction at low energies. Using the vertices
contained in (\ref{pi-di}) one can study pion-nucleon scattering at small
pion momenta provided we have a good model description of bounding diquark
together with the third quark into nucleon.
Of cause, in order to obtain the predictions comparable with the data the
influence of quark masses should be taken into account. However it is
important that we succeeded in description of pion-diquark interaction
from universal point of view of extended chiral transformation approach.

\section*{Acknowledgments}

The author is very indebted to Prof. Yu.V. Novozhilov and
Dr. D.V. Vassilevich for many interesting and stimulating
discussions.

\end{document}